\documentclass[aps,preprint]{revtex4-1}
\usepackage{latexsym}
\usepackage{graphicx}
\usepackage[dvipsnames]{xcolor}
\usepackage{xspace}
\usepackage{amsmath}
\usepackage{amsfonts}
\usepackage{amssymb}
\usepackage{bm}
\usepackage[colorlinks=true,linkcolor=blue,citecolor=blue,pdfauthor=Gross]{hyperref}

\begin{document}
\title{Advances in Single Crystal Growth and Annealing Treatment of Electron-doped HTSC}

 \author{Michael Lambacher,
   Toni Helm, 
    Mark Kartsovnik,
    and Andreas Erb\thanks{electronic mail: Andreas.Erb@wmi.badw.de}
}

\affiliation{Walther-Mei{\ss}ner-Institut, Bayerische Akademie der Wissenschaften,
           Walther-Mei{\ss}ner-Stra{\ss}e 8, D-85748 Garching, Germany.
 }

\begin{abstract}
High quality electron-doped HTSC single crystals of $\rm
Pr_{2-x}Ce_{x}CuO_{4+\delta}$ and $\rm Nd_{2-x}Ce_{x}CuO_{4+\delta}$ have been
successfully grown by the container-free traveling solvent floating zone 
technique. The optimally doped $\rm Pr_{2-x}Ce_{x}CuO_{4+\delta}$ and $\rm
Nd_{2-x}Ce_{x}CuO_{4+\delta}$ crystals have transition temperatures $T_{\rm c}$
of $25$\,K and $23.5$\,K, respectively, with a transition width of less than
$1$\,K. We found a strong dependence of the optimal growth parameters on the Ce
content $x$. We discuss the optimization of the post-growth annealing treatment
of the samples, the doping extension of the superconducting dome for both
compounds as well as the role of excess oxygen. The absolute oxygen content of
the as-grown crystals is determined from thermogravimetric experiments and is
found to be $\ge 4.0$. This oxygen surplus is nearly completely removed by a
post-growth annealing treatment. The reduction process is reversible as
demonstrated by magnetization measurements. In as-grown samples the excess
oxygen resides on the apical site O(3). This apical oxygen has nearly no doping
effect, but rather influences the evolution of superconductivity by inducing
additional disorder in the CuO$_{2}$ layers. The very high crystal quality of
$\rm Nd_{2-x}Ce_{x}CuO_{4+\delta}$ is particularly manifest in magnetic quantum
oscillations observed on several samples at different doping levels. They
provide a unique opportunity of studying the Fermi surface and its dependence
on the carrier concentration in the bulk of the crystals.
\end{abstract}

 \maketitle

\section{Introduction}
\label{sec:intro}

In comparison to the various families of the hole-doped cuprate superconductors
the electron-doped 214 compounds Ln$_{2-x}$$\rm Ce_xCuO_{4+\delta}$ (with Ln$=$
Pr, Nd, Sm, Eu,$\ldots$) have been studied less intensively, even though this
system is known for more than 15 years~\cite{tokura89}. There are different
reasons for this. First, the growth of high quality single crystals and the
post-growth annealing treatment are very demanding. Second, the controlled and
reproducible preparation of well characterized samples for different
experiments, such as well-oriented samples with clean surfaces for optical
spectroscopies, is difficult. Despite these drawbacks, the electron-doped
$214$-compounds are very attractive as a sample set for systematic
investigations for two reasons: (i) They are solid solutions with a simple
tetragonal crystal structure and nearly the whole phase diagram can be probed
using only a single compound. Starting from the undoped antiferromagnetic
insulator Ln$_{2}$$\rm CuO_{4+\delta}$ (with Ln $=$ Pr, Nd, Sm, Eu,$\ldots$),
samples with different electronic ground states up to the overdoped metallic
regime can be grown by gradually doping with tetravalent Ce. (ii) With $\rm
La_{2-x}Sr_xCuO_{4-\delta}$ there is a corresponding hole-doped compound with a
similar crystal structure and critical temperature $T_{\rm c}$. This provides
the unique opportunity to analyze similarities and differences in the phase
diagram of the electron- and hole-doped compounds.

\section{Growth Process and Growth Parameters}
\label{sec:2}

\subsection{Crystal Growth}
\label{sec:20}

The growth of high quality $214$-crystals in crucibles from a CuO-rich melt is
difficult for various reasons: Firstly, deviations of the distribution
coefficient from unity in these systems give rise to gradients in the dopant
concentration of the grown crystals. Secondly, it is well known from other
cuprate systems, such as $\rm YBa_{2}Cu_{3}O_{7-\delta}$~\cite{Erb}, as well as
from previous work on $\rm Nd_{2-x}Ce_{x}CuO_{4+\delta}$~\cite{maljuk00}, that
the aggressive melts partly dissolve the most common available crucible
materials. This unwanted corrosion leads to impurities and, in turn, to a
deterioration of the crystal quality. We also note that the $214$-compounds are
incongruently melting solid solutions~\cite{oka89}. Therefore, the growth
process is restricted to a small interval between the peritectic and eutectic
points of the compositional phase diagram.

The Traveling Solvent Floating Zone (TSFZ) technique provides the opportunity
and flexibility to work at a certain favorable working point within this small
growth window by using a small CuO-rich flux pellet and a corresponding
atmosphere and temperature. The flux pellet forms a vertical local melt, held
by surface tension, where the polycrystalline feed material is dissolved at the
top and re-crystallized at the bottom on a seed. Using suitable growth
conditions, an equilibrium between growth and solubility rate can be obtained.
In this way large crystals of several centimeters length can be grown under
accurately controllable stable conditions (flux composition, temperature,
oxygen partial pressure). This is the basic prerequisite for homogeneous
crystals.

For the growth of a series of Ln$_{2-x}$$\rm Ce_{x}CuO_{4+\delta}$ crystals
(with Ln $=$ Nd, Pr) we have used a $4$-mirror furnace with four $300$\,W
halogen lamps. The polycrystalline feed rods and flux pellets were prepared in
the following way: For the feed rods the corresponding rare earth oxides and
CuO (with purity of $99.99\%$) were mixed according to the desired
stoichiometric composition of the $214$-compounds. Then, phase pure ceramic
samples were obtained by a five-fold pre-reaction of the powders at
temperatures of $900^\circ$C, $920^\circ$C, $950^\circ$C, and two times at
$980^\circ$C for $10$\,h in air. Between each calcination step the pre-reacted
powder was homogenized by grinding. The multiple calcination promotes the
homogeneity. The phase formation of the powder was checked by X-ray powder
diffraction. The phase pure ceramic samples were pressed hydrostatically to
rods of $7$ mm in diameter and $140$ mm in length. These polycrystalline rods
were sintered in air at temperatures of $1050^\circ$C, $1100^\circ$C, and
$1200^\circ$C for $5$\,h, respectively. These high temperatures close to the
peritectic temperature are important in order to increase the density, avoiding
the absorption of the liquid flux during crystal growth. The flux material with
the composition ratio $[(2-x)\cdot 1/2\cdot ($Ln$_{2}\mathrm{O}_{3})+ x\cdot
\mathrm{CeO}_{2}]/\mathrm{CuO} = 15/85$ was pre-reacted in the same way as the
feed rods, followed by an annealing step at $1010^\circ$C for $10$\,h in air.

For single crystal growth, a piece of a polycrystalline rod was used as a seed.
On this seed a flux pellet of $0.35-0.40$\,g was placed. The mass and size of
this flux pellet needs to be adjusted to the dimensions of the radiation focus
and the diameter of the rods. The vertical molten zone has a diameter and
length of $5-6$\,mm and can easily be held by surface tension during the entire
growth process. The growth atmosphere, i.e. the oxygen partial pressure $p_{\rm
O_{2}}$, was found to be the most critical parameter. It affects both the
growth temperature and the crystallizing phases, as well as the stability of
the growth process. While undoped compounds Ln$_{2}$$\rm CuO_{4+\delta}$ grow
stable in a pure O$_{2}$ atmosphere at a pressure of $4-5$\,bar, the growth at
low $p_{\rm O_{2}}$ is unstable. The opposite situation was found for doped
Ln$_{2-x}$$\rm Ce_{x}CuO_{4+\delta}$, even for relatively small Ce content $x$.
Our growth experiments clearly showed that with increasing Ce content $x$ a
decreasing oxygen partial pressure $p_{\rm O_{2}}$ is required for the optimal
growth conditions.

The set of $\rm Nd_{2-x}Ce_{x}CuO_{4+\delta}$ crystals was grown in a mixture
of Ar/O$_{2}$ with $p_{\rm O_{2}} = 0.20$\,bar for $x\leq 0.15$ and $p_{\rm
O_{2}} \leq 0.15$\,bar for $x\geq 0.15$. The set of $\rm
Pr_{2-x}Ce_{x}CuO_{4+\delta}$ crystals was also grown in an Ar/O$_{2}$
atmosphere with $p_{\rm O_{2}} = 0.20$\,bar for $x\leq 0.08$ and $p_{\rm O_{2}}
= 0.03$\,bar for $x\ge 0.08$. In order to suppress Cu-evaporation from the
melt, a pressure of $3-4$\,bar was applied. As the evaporation of Cu is very
small compared to the used amount of flux, this loss can be balanced by the
flux itself without changing the growth conditions. We note that the change of
the flux composition due to the CuO evaporation can be estimated from the
weight loss of the whole system before and after the growth process and does
not exceed 2\%. The evaporation in $\rm Nd_{2-x}Ce_{x}CuO_{4+\delta}$ is
slightly higher than in the $\rm Pr_{2-x}Ce_{x}CuO_{4+\delta}$ system. A
surplus of $1-2$\% CuO in the feed rod composition for compensating this
evaporation makes it difficult to control the growth process, which often
results in CuO inclusions and (during the long lasting growth experiments)
unstable solvent zones. Apart from the CuO inclusions, precipitations of
CeO$_{2}$ on microscopic scale and macroscopic clustering of the dopant can
only be avoided by means of an appropriate atmosphere with low $p_{\rm O_{2}}$.
Doped crystals grown in pure O$_{2}$, as it is often suggested in
literature~\cite{kurahashi02,mang04,uefuji03}, or at too high oxygen partial
pressure for the respective Ce content $x$ are inhomogeneous. This results in
broad transition curves as well as in ambiguous and varying relations between
$T_{\rm c}$ values and the nominal Ce content $x$. Moreover, a slow
decomposition of the compounds over longer time periods (usually several months
to several years) is obtained. In order to support the homogeneity of the
molten zone, the seed and feed rods are rotated in opposite directions at about
$20$\,rpm. The growth rate for all crystals was $0.5$\,mm/h as determined by
the external movement of the mirror system. These small growth rates and also
small diameters of the grown rods proved to be important for obtaining
rods consisting of only a single grain. We note that thinner diameters of the
crystallized rod can be obtained by pulling away the feed rod simultaneously
during the growth process.

\begin{figure}[tb]
  \centering
  \includegraphics[width=.95\columnwidth]{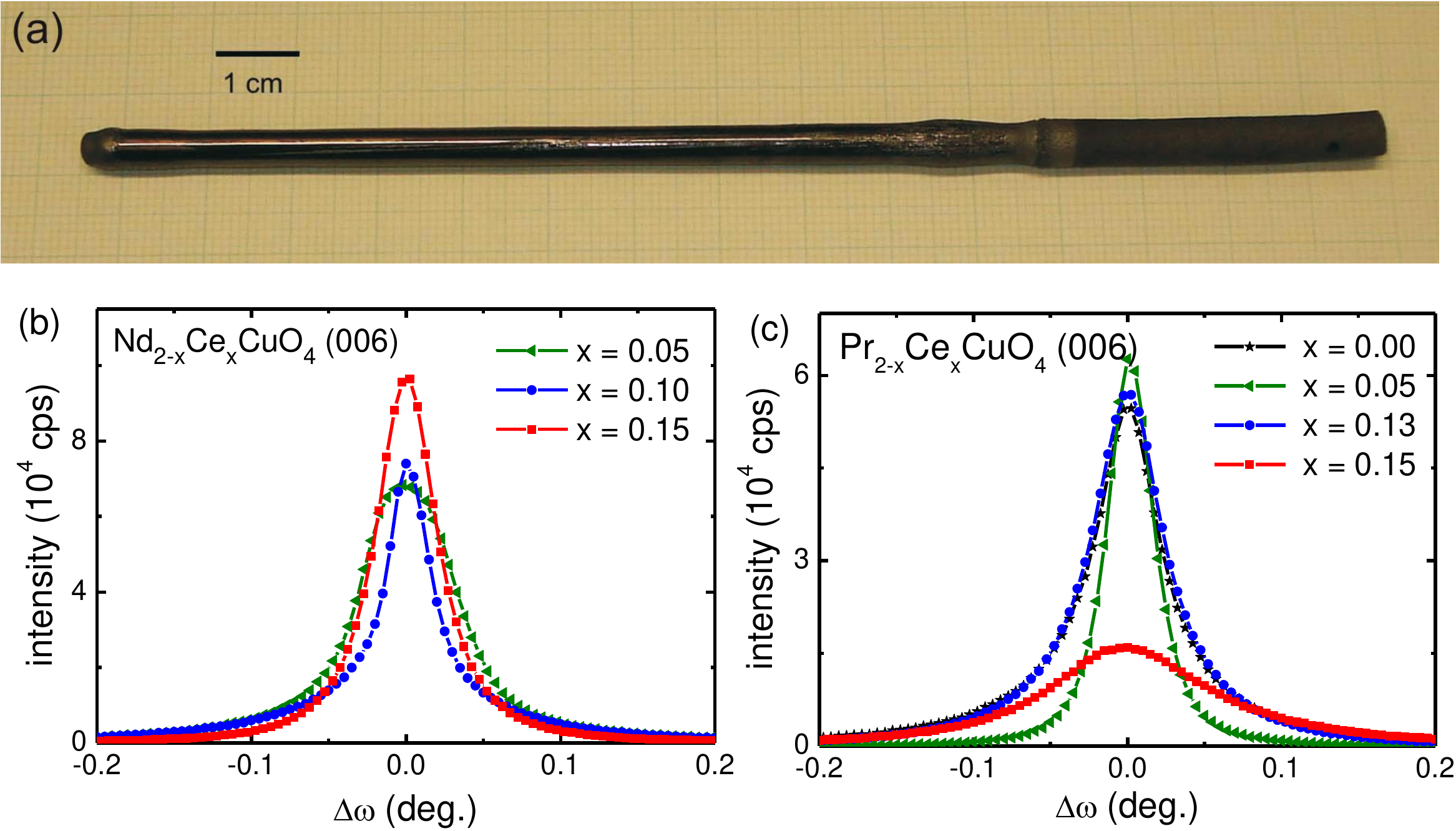}
  \caption{
(a) Picture of a TSFZ-grown $\rm Nd_{1.85}Ce_{0.15}CuO_{4+\delta}$ crystal. (b)
and (c) show rocking curves of the $(006)$ peak of
$\rm Nd_{2-x}Ce_{x}CuO_{4+\delta}$ and  $\rm Pr_{2-x}Ce_{x}CuO_{4+\delta}$
crystals, respectively. The full width at half maximum (FWHM) for all crystals
with different Ce contents is less than $0.08^\circ$.}
 \label{fig1}
\end{figure}

For all crystals the preferred growth direction was along the crystallographic
$[110]$ or $[100]$ axis. This is caused by the fact that the grains grow much
faster along the CuO-layers than perpendicular to them. In this way only grains
with the favorable orientation survive during the growth process. The
possibility of the simultaneous growth of several grains with slightly
different orientations restricts the use of the samples in experiments, where
all these grains have to be extracted and individually oriented. This is for
example the case in neutron scattering experiments, where big crystals of
several grams are required. A way out would be to influence the growth
direction by using a single crystal as seed. However, this has been found
difficult, as marginal deviations from the optimal growth conditions at the
beginning of the growth process lead to the nucleation of new grains or the
crystallization of small amounts of flux material. Fig.~\ref{fig1}(a) shows a
grown crystal rod of $\rm Nd_{1.85}Ce_{0.15}CuO_{4+\delta}$ with a $c$-oriented
facet, which is present on the last $\sim 2$\,cm of the rod [left-hand side in 
Fig.~\ref{fig1}(a)] .

\subsection{Crystal Characterization}
\label{subsec:21}

The metal composition along and perpendicular to the growth direction of the
crystals was controlled by energy dispersive X-ray analysis (EDX). Within
the accuracy of the measurement the grown single crystals exhibit the same
chemical composition as the feed rods: no inhomogeneities or gradients of the
rare earth concentrations in growth as well as in the radial direction were found.
Small deviations in the doping concentration ($x=0.01$), microscopic
precipitations of $\rm Ln_{2}$O$_{3}$ as well as variations in the oxygen
distribution within the specimens could not be detected by EDX. More accurate
information about the microscopic homogeneity can be obtained from
magnetization measurements and from the onset and the shape of
superconducting transition curves of different crystals, which are annealed at
well defined conditions. The solubility limit of Ce in $\rm
Nd_{2-x}Ce_{x}CuO_{4+\delta}$ and $\rm Pr_{2-x}Ce_{x}CuO_{4+\delta}$ is found
at $x=0.18$ and $x=0.15$, respectively. X-ray analysis confirms the phase
purity and the T'-structure of the as-grown crystals. Moreover, the rocking
curves of the (006)-peak show a full width at half maximum FWHM $\le 0.08$ for
all doping levels [see Fig.~\ref{fig1}(b)]. These values are --to our
knowledge-- the smallest values reported so far for electron-doped
$214$-compounds. The mosaicity of the crystals is hardly improved by the
annealing treatment. However, we can confirm the formation of small amounts of
epitaxially grown (Ln,Ce)$_{2}$O$_{3}$ impurity
phases~\cite{matsuura03,mang04,kang05} due to the annealing treatment, which is
independent from the doping concentration. Rods consisting of two or several
grains are separated and oriented before annealing. Bulk superconductivity was
verified by specific heat measurements. We emphasize that pieces cut from the
inner part of large crystals have the same transition curves ($T_{\rm c}$,
$\Delta T_{\rm c}$) as the whole crystal itself. Thus, core-shell effects can
be ruled out. The as-grown and annealed crystals were stable over several
years, without changing their physical properties. Crack- and inclusion-free
high quality crystals are difficult to cleave. Therefore, for measurements
(spectroscopic measurements, transport, etc.) the crystals had to be oriented,
cut, and polished in an appropriate way. This time consuming work is a drawback
compared to the easily cleaving compounds of the Bi-family and $\rm
YBa_{2}Cu_{3}O_{7-\delta}$ which grow in a platelet shape.

\section{Annealing Treatment}
\label{sec:3}

\subsection{Annealing Procedure}
\label{sec:30}

An annealing treatment of the crystals after the growth process eliminates
tension in the crystal and disorder in the metal sublattice. At the same time
it removes interstitial oxygen. The as-grown $\rm Pr_{2-x}Ce_{x}CuO_{4+\delta}$
crystals, grown in an atmosphere of $p_{\rm O_{2}} = 0.03$\,bar, are not
superconducting, whereas $\rm Nd_{2-x}Ce_{x}CuO_{4+\delta}$, grown at the
same $p_{\rm O_{2}}$, shows a broad transition with $T_{\rm c} = 10$\,K.
Unfortunately, the growth process at such low $p_{\rm O_{2}}$ for $\rm
Nd_{2-x}Ce_{x}CuO_{4+\delta}$ is not stable enough. Thus, growth atmospheres
with higher oxygen partial pressure as described above were used. This results
in excess oxygen in the as-grown samples requiring a post-growth annealing
treatment.

\begin{figure}[tb]
  \centering
  \includegraphics[width=.65\columnwidth]{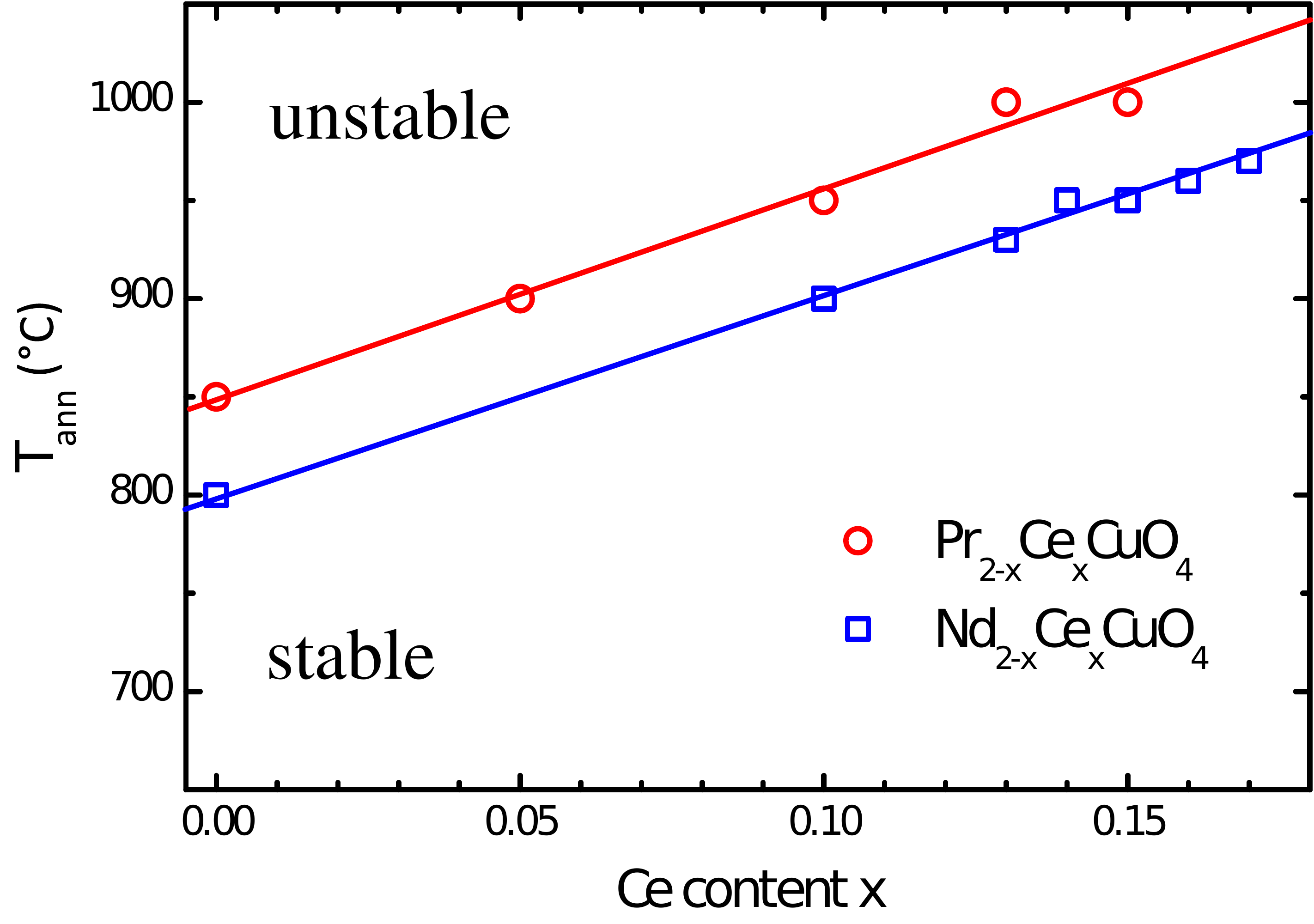}
  \caption{
Doping dependence of the maximum annealing temperature $T_{\rm ann}$ for
$\rm Nd_{2-x}Ce_{x}CuO_{4+\delta}$ and $\rm Pr_{2-x}Ce_{x}CuO_{4+\delta}$.
The lines are guides to the eye. All samples are annealed in a flow of pure Ar
$4.8$ (O$_{2}$ $\le 3$\,ppm) for $t_{\rm ann} =20$\,h.}
 \label{fig2}
\end{figure}

In the past, many studies concerning the optimal annealing parameters, phase
decomposition and oxygen loss have been carried
out~\cite{mang04,kurahashi02,shibata01,kim93}. The general consensus of all
these studies is, that bulk superconductivity and sharp transition curves are
only realized by a severe reduction treatment of the as-grown samples. In this
process a small doping dependent amount of the excess oxygen $\delta = 0.02 -
0.06$ is removed. Nevertheless, fundamental questions regarding the onset,
shape and doping interval of the superconducting dome, the absolute value of
the oxygen content of the as-grown and reduced samples, the role of additional
oxygen in the compound, and the reversibility of the reduction step have not
yet been settled and currently are controversially discussed. In order to verify 
the different results, we annealed the samples in the following way: As-grown
single crystals were annealed for $t_{\rm ann} =20$\,h in a flow of pure Ar
$4.8$ (O$_{2}$ $\le 3$\,ppm) at constant temperatures close to the stability
limit. The stability limit for this annealing treatment depends on the Ce
content as well as on the rare earth element $Ln$ and is shown in
Fig.~\ref{fig2}.

\begin{figure}[tb]
  \centering
  \includegraphics[width=.95\columnwidth]{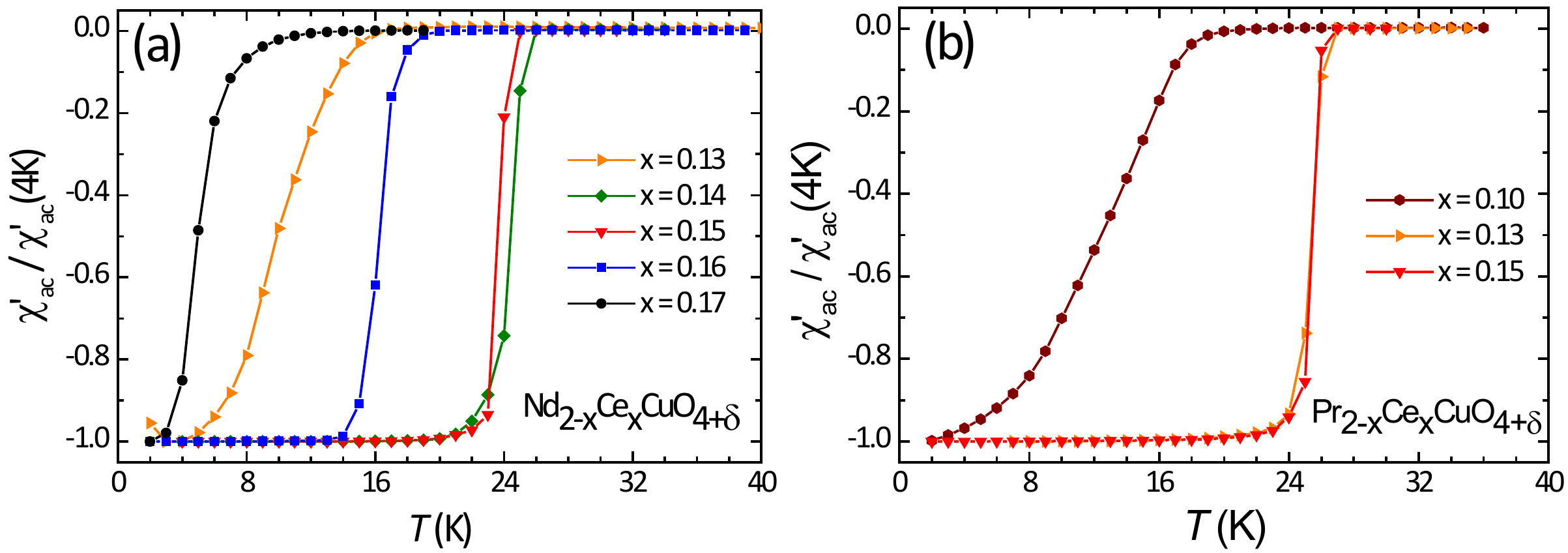}
  \caption{
Ac-susceptibility of $\rm Nd_{2-x}Ce_{x}CuO_{4+\delta}$ (a) and
$\rm Pr_{2-x}Ce_{x}CuO_{4+\delta}$ (b) single crystals plotted versus
temperature for samples with different Ce contents. For a better comparison the
measured diamagnetic susceptibilities are normalized to their low temperature
values. Note, that the sample mass ranges from $100$ to $250$\,mg. For
$\rm Nd_{1.85}Ce_{0.15}CuO_{4+\delta}$ we obtain $T_{\rm c}=23.6$\,K and
$\Delta T_{\rm c}=1.3$\,K, whereas for $\rm Pr_{1.85}Ce_{0.15}CuO_{4+\delta}$
we find  $T_{\rm c}=25.5$\,K and $\Delta T_{\rm c}=1.4$\,K.}
 \label{fig3}
\end{figure}

For the annealing step, the crystals were enclosed in polycrystalline crucibles
of the same material in order to provide a homogeneous environment and to
protect the crystal surface. After the treatment at high temperatures the
crystals were cooled down to room temperature at moderate cooling rates. The
crystals were not quenched, since the intention was to remove the excess oxygen
but not to freeze-in a specific O$_{2}$ ordering state. In this way we 
avoided additional tension in the metallic sublattice which could occur upon
quenching. This procedure was applied to all doped crystals of both compounds.
Depending on the Ce content $x$, the crystals exhibit sharp transition curves
as shown in Fig.~\ref{fig3}. It has been often reported in
literature~\cite{mang04,kurahashi02} that the annealing treatment depends on
the sample size. Moreover, it was suggested that a subsequent additional
annealing step after the reduction step, which is then performed in pure
O$_{2}$ at moderate temperatures of $500 - 600^\circ$C for $10$\,h increases
$T_{\rm c}$. However, for our crystals with masses ranging from $50$ to
$250$\,mg we do not see any size effect of the transition curves. This
observation confirms that the oxygen diffusion coefficient $D(T)$ and diffusion
lengths $l \sim \sqrt{D(T) \cdot t_{\rm ann}}$ are large enough at the
annealing temperatures used in our experiments. Moreover, a second, short
annealing treatment at moderate temperatures in pure O$_{2}$ complicates the
control of the oxygen content and the microscopic homogeneity of the crystal,
as $D(T)$ might be completely different at the different annealing steps at
high and moderate temperatures. For example, it is well known from diffusion
studies on YBa$_{2}$Cu$_{3}$O$_{7-\delta}$ and RE-$123$ single crystals, that
$D(T)$ varies over several orders of magnitude. Since there are no reliable
studies of the kinematics of the oxygen diffusion process for the
$214$-crystals, results of annealing experiments under different atmospheres,
temperatures and time scales are speculative. We also note that studies on
large ceramic samples give misleading results, as the in- and out-diffusion of
oxygen is mainly governed by grain boundaries, micro-cracks and other
imperfections of the specimen.

\subsection{Absolute Value of the Oxygen Concentration}
\label{sec:31}

Thermogravimetric experiments and subsequent x-ray powder diffraction on high
quality as-grown single crystals are performed in order to estimate the
absolute oxygen content $4+\delta$ per formula unit. The absolute oxygen
concentration is required for an accurate determination of the oxygen loss
during the annealing treatment from the measured mass difference between
as-grown and annealed samples. As the oxygen non-stoichiometry in the
$214$-compounds is very small, only crack- and inclusion-free single crystals
of $100 - 150$\,mg mass were used for these experiments in order to avoid
misleading results. Systematic errors coming from the experimental setup are
minimized by recording a base line ahead of each measurement. The oxygen
concentration is calculated from the measured weight loss during the reduction
in a flow of forming gas with composition $\mathrm{H}_{2}/\mathrm{N}_{2}=15/85$
for $1$\,h at $1000^\circ$C. The crystals decompose according to the following
equation, where $Ln =$Nd,~Pr:
\begin{eqnarray*}
2\mathrm{Ln}_{2-x}\mathrm{Ce}_{x}\mathrm{CuO}_{4+\delta}  & \rightarrow & (2-x) \mathrm{Ln}_{2}\mathrm{O}_{3} +
2x\;\mathrm{CeO} _{2} + 2\mathrm{Cu^{ele}} + \left( 1+\delta - \frac{x}{2} \right) 
\mathrm{O}_{2}\;.
\end{eqnarray*}
Undoped as-grown crystals show a nearly stoichiometric oxygen content. Upon
increasing the Ce content $x$ the oxygen surplus $\delta$ increases slightly
and has values of about $0.03$ for optimally doped $\rm
Nd_{1.85}Ce_{0.15}CuO_{4+\delta}$ and $\rm Pr_{1.85}Ce_{0.15}CuO_{4+\delta}$.
Note, that the absolute value depends on the oxygen partial pressure $p_{\rm
O_{2}}$ used during the crystal growth. This might also be an explanation of
different values in literature reported for the oxygen non-stoichiometry and
the oxygen loss during annealing, as the crystals are grown in different
atmospheres. For crystals with Ce concentrations $x$ within the superconducting
dome the oxygen loss amounts to $0.02$ for $\rm Nd_{2-x}Ce_{x}CuO_{4+\delta}$
and $\rm Pr_{2-x}Ce_{x}CuO_{4+\delta}$. Thus, the annealed samples have a
stoichiometric number of $4+\delta = 4.01$ for oxygen, which hardly differs
from the nominal value of $4.0$ within the uncertainty of the experiment.
Therefore, as discussed in more detail below, oxygen does not play an important
role regarding the doping level and we can set $n =x$. Here, $n$ is the carrier
concentration per Cu-ion and $x$ is the Ce concentration.

\section{Doping Dependent Phase Diagram}
\label{sec:4}

\subsection{The Superconducting Dome}
\label{sec:40}

\begin{figure}[tb]
  \centering
  \includegraphics[width=.95\columnwidth]{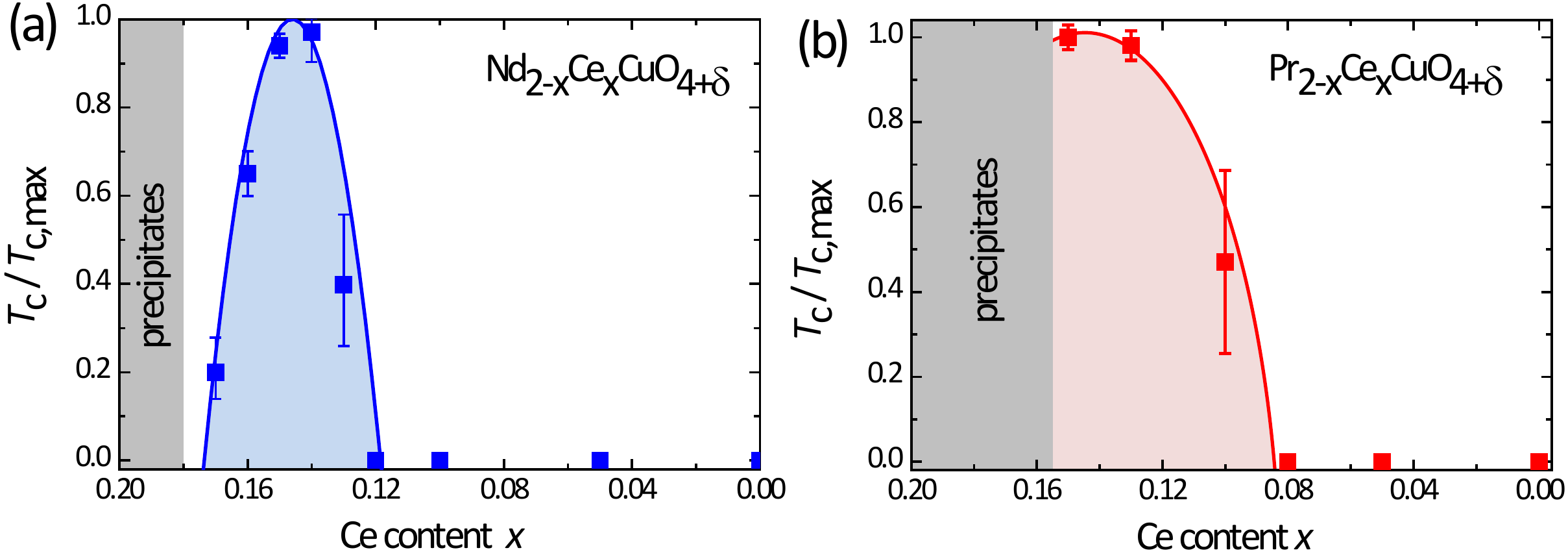}
  \caption{
Normalized transition temperature of $\rm Nd_{2-x}Ce_{x}CuO_{4+\delta}$ (a)
and $\rm Pr_{2-x}Ce_{x}CuO_{4+\delta}$ (b) single crystals plotted versus
the Ce concentration $x$. The $T_{\rm c}$ values have been derived from the
transition curves in Fig.~\ref{fig3}. The data points mark the $T_{\rm c}$
values where the resistivity dropped to $50$\% of its normal state value. The
vertical bars indicated the transition widths $\Delta T_{\rm c}$ in which the
resistivity dropped from 90\% to 10\%. In (a) the experimental data of the
overdoped regime are fitted by the quadratic function resulting in the
empirical relation (\ref{eq:Tcx}). The superconducting dome is asymmetric with
an abrupt onset at $x=0.13$, an expected maximum at $x=0.146$ and a steady
decrease towards zero before the solubility limit at $x=0.18$ is reached. For
the $\rm Pr_{2-x}Ce_{x}CuO_{4+\delta}$ system shown in (b) the overdoped regime
cannot be probed due to precipitations. The optimal doping is found at $x=0.15$
and the underdoped region extends to $x=0.10$, where superconductivity abruptly
sets in. }
 \label{fig4}
\end{figure}

Fig.~\ref{fig4} shows the transition temperatures as a function of the Ce
content $x$ for both electron-doped $214$-compounds discussed in this paper. We
recall that in the hole-doped cuprates superconductivity appears
within a broad doping interval of $p=[0.05,0.27]$ and $T_{\rm c} \propto
(p-p_{\rm opt})^2$, where $p_{\rm opt}$ is the optimum doping level 
corresponding to the maximum $T_{\rm c}$ value~\cite{tallon95}. In contrast,
for the electron-doped compounds $\rm Nd_{2-x}Ce_{x}CuO_{4+\delta}$ and $\rm
Pr_{2-x}Ce_{x}CuO_{4+\delta}$ the superconducting dome is restricted to a much
smaller doping range and does in general not show a quadratic dependence
$T_{\rm c} \propto (n-n_{\rm opt})^2$. As shown in Fig.~\ref{fig4}, for the
$\rm Nd_{2-x}Ce_{x}CuO_{4+\delta}$ system superconductivity abruptly sets in at
$n\simeq x=0.13$ and $T_{\rm c}$ jumps up to $T_{\rm c,max}$ $n\simeq x=0.146$.
In the overdoped region the $T_{\rm c}(x)$ dependence can be well described by
the empirical quadratic function (solid line in Fig.~\ref{fig4})
\begin{eqnarray}
\frac{T_{\rm c}}{T_{\rm c,max}} & = & 1 - 1320 \; (x-0.146)^2
\label{eq:Tcx}
\end{eqnarray}
with $T_{\rm c,max}=25.1$\,K. According to this empirical curve, $T_{\rm c}$
becomes zero at $x=0.118$ and $x=0.173$. A comparison of the empirical curve
with the experimental results in the underdoped region reveals a slight
asymmetry of the superconducting dome in the $\rm Nd_{2-x}Ce_{x}CuO_{4+\delta}$
system. However, more crystals with even finer variations in the Ce
concentration in the range $[0.12, 0.13]$ are required to clarify the onset
point and the sudden occurrence of superconductivity with $T_{\rm c}$ values of
about $0.4\;T_{\rm c,max}$ at $x=0.13$. From the doping point of view the
overdoped region can be controlled in a better way. Notably, the whole
superconducting region down to $T_{\rm c} =0$ can be probed, as a solubility
limit of $x=0.18$ is found for Ce in the $\rm Nd_{2-x}Ce_{x}CuO_{4+\delta}$
system. In this context the width of the transition curves as a measure of the
quality of the crystals has to be discussed. It is clear, that crystals near
optimal doping show the sharpest transition curves, since slight variations in
the doping of the specimen do not strongly affect $T_{\rm c}$. In the overdoped
region, the crystals are very sensitive to small doping variations due to the
steep slope of the $T_{\rm c}(x)$ curve defining the superconducting dome. For
example, for a sample with nominal doping $x=0.17$ and a spatial variation in
the doping level as small as $\Delta x =0.002$, a transition width of $\Delta
\ge 3$\,K is expected. In the underdoped region the situation is even more
complicated and it is not clear whether inhomogeneities of the crystals or
other effects are responsible for the observed broad transition curves.

In the $\rm Pr_{2-x}Ce_{x}CuO_{4+\delta}$ system, the overdoped region cannot
be reliably probed due to precipitations of the dopant. Therefore, the
comparison of the superconducting phase diagram is restricted to the optimal
and underdoped regions. Comparing the $\rm Pr_{2-x}Ce_{x}CuO_{4+\delta}$ and
$\rm Nd_{2-x}Ce_{x}CuO_{4+\delta}$ system, we have to point to two common
properties: first, the optimum doping level $x=0.15$ is about the same and,
second, both systems show an abrupt onset of superconductivity in the
underdoped region. For $\rm Pr_{2-x}Ce_{x}CuO_{4+\delta}$, this onset is found
at $x=0.10$. However, there are also pronounced differences in the $T_{\rm
c}(x)$ dependence of both systems. It is evident from Fig.~\ref{fig4} that the
$\rm Pr_{2-x}Ce_{x}CuO_{4+\delta}$ system shows a much broader $T_{\rm c}(x)$
dependence in the underdoped regime. This difference in the shape of the
superconducting dome is most likely the result of material specific issues and
is a general problem when comparing electron-doped compounds. Therefore, in the
future one main task should be the elimination of material specific factors 
in order to merge the different phase diagrams of the electron-doped compound
into a general one.

\subsection{Role of Oxygen }
\label{sec:41}

\begin{figure}[tb]
  \centering
  \includegraphics[width=.95\columnwidth]{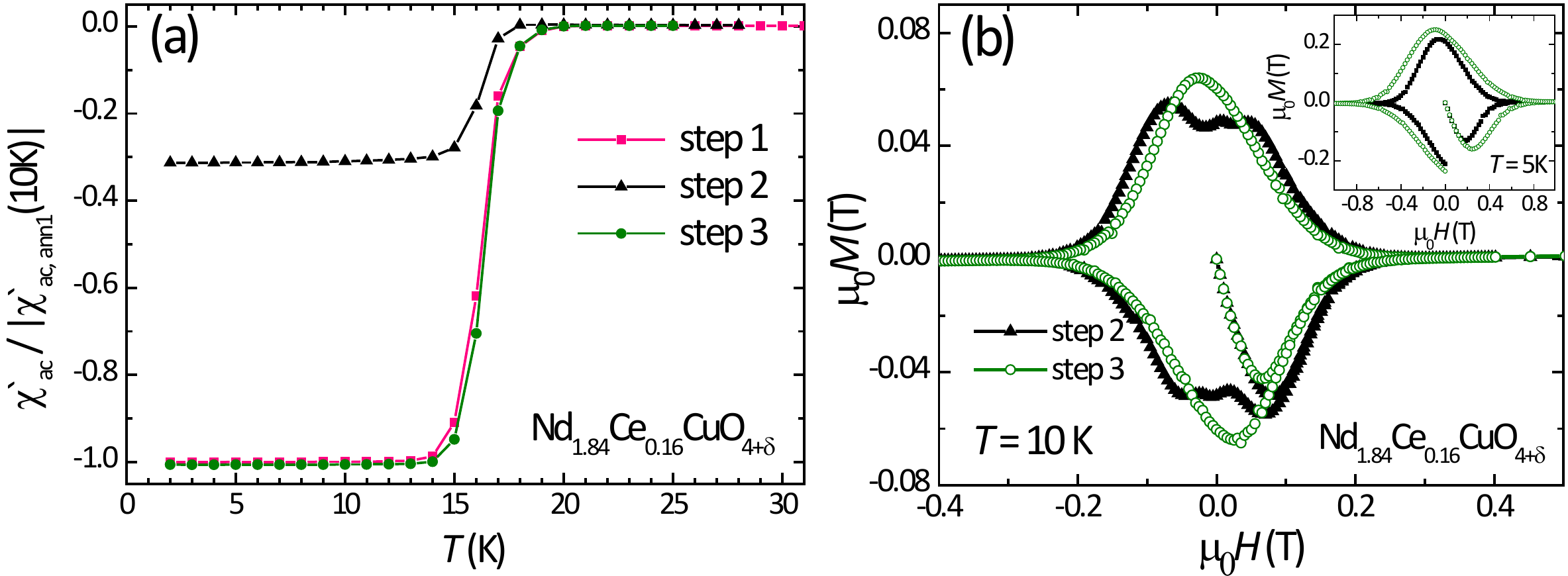}
  \caption{
(a) Temperature dependence of the normalized ac susceptibility recorded for a
$c$-oriented $\rm Nd_{1.84}Ce_{0.16}CuO_{4+\delta}$ crystal of $51.5$\,mg after
different steps of an annealing experiments. Step~1: Reduction of the as-grown
sample in flowing of Ar($4.8$) at $960^\circ$C for $20$\,h
($\Delta\delta=-0.018$). Step~2: Oxygenation of the reduced sample in a flowing
O$_2$/Ar$=0.1\%$  at $750^\circ$C for $70$\,h ($\Delta\delta=+0.005$). Step~3:
Repetition of step~1 with the same parameters ($\Delta\delta=-0.006$). The
ac-susceptibility curves confirm the reversibility of the annealing steps. The
transition curve after oxygenation shows $T_{\rm c}$ reduced by 1\,K
and a smaller (by a factor of 3) diamagnetic susceptibility. (b) Magnetization
curves recorded at $5$\,K (inset) and $10$\,K after the annealing steps~2 and
3. After partial oxygenation the crystal shows a broader magnetization curve
with irregularities. This fluctuating fishtail effect, which appears/disappears
with oxygenation/reduction is a clear signature for oxygen induced disorder,
which gradually destroys superconductivity. }
 \label{fig5}
\end{figure}

The precise experimental control of the oxygen concentration is one of the main
difficulties when preparing 214-compounds. It is challenging to tune and
determine the small oxygen variations in a crystal in a non-destructive manner.
Therefore, many and partly contradicting statements concerning the oxygen
occupation of the different sites in the crystal structure of as-grown and
annealed crystals exist~\cite{higgins06,radaelli94,richard04}. Oxygen is often
believed to act as co-dopant in such reports. In order to clarify whether or
not oxygen non-stoichiometry may act as a second doping channel, an overdoped
$\rm Nd_{1.84}Ce_{0.16}CuO_{4+\delta}$ high quality single crystal was used for
oxygenation experiments. Strongly overdoped crystals are expected to be
extremely sensitive to oxygenation experiments, because of the steep slope of
the superconducting dome (cf. Fig.~\ref{fig4}) in this regime.

\begin{figure}[tb]
  \centering
  \includegraphics[width=.95\columnwidth]{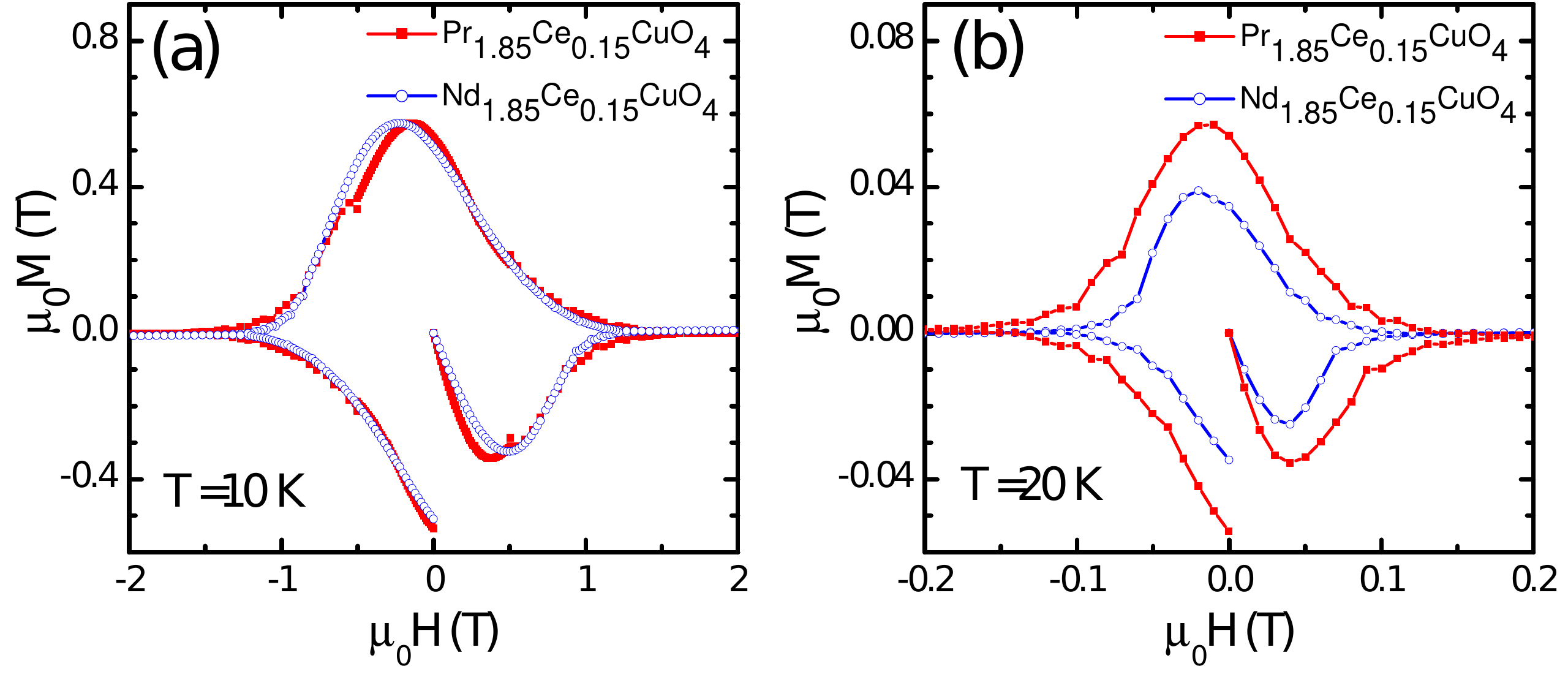}
  \caption{
Magnetization curves of a $c$-oriented (a)
Nd$_{1.85}$Ce$_{0.15}$CuO$_{4+\delta}$ ($182$\,mg) and (b)
Pr$_{1.85}$Ce$_{0.15}$CuO$_{4+\delta}$ ($100$\,mg) single crystal. The
corresponding transition curves are shown in Fig.~\ref{fig3}. No anomalies are
detected, thereby confirming the microscopic homogeneity of the crystals. }
 \label{fig6}
\end{figure}

In a simple ionic model of the electron-doped 214-compounds the oxygen
contribution to the electron concentration $n$ per Cu-ion is given by the
relation
\begin{eqnarray*}
n & = & x-2\delta \; .
\end{eqnarray*}
Here, the valences for the constituents are assumed to be $\mathrm{Ce}=4+,
\mathrm{Ln}=3+, \mathrm{O}=2-, \mathrm{Cu}=\nu$. The formal valency of Cu is $\nu =
2-n=2-x+2\delta$. In this picture the presence of any excess oxygen with amount
$\delta$ lowers the charge carrier concentration $n$ and, hence, compensates
the carrier concentration induced by the Ce content $x$. By starting with a
reduced overdoped $\rm Nd_{1.84}Ce_{0.16}CuO_{4.01}$ crystal, the optimally
doped and underdoped regions of the superconducting dome should be reached
easily by appropriate oxygenation. In Fig.~\ref{fig5}(a) the change of the
transition curve by an oxygen variation of $\Delta\delta=+0.005$ is presented.
Clearly, $T_{\rm c}$ is lowered by $1$\,K and the absolute value of the
ac-response at $2$ K is reduced by a factor of about 3. This is in complete
contradiction to our simple model and the measured phase diagram. According to
$n = x-2\delta$, for the $\rm Nd_{1.84}Ce_{0.16}CuO_{4.01}$ sample the carrier
concentration $n$ is expected to change from $n=0.14$ to $n=0.13$ on going from
O$_{4.010}$ to O$_{4.015}$ and, thus, an increase of $T_{\rm c}$ towards
$T_{\rm c,max}$ is expected. A subsequent standard reduction step removes the
inserted oxygen again ($\Delta\delta=-0.006$) and the crystal shows the same
sharp transition curve as before oxygenation. From this annealing experiment we
arrive at two conclusions: First, the metal sublattice remains unaffected by
carefully performed oxygenation experiments. In particular, no irreversible
microscopic CeO$_2$-precipitations or rearrangements occur, which could be
responsible for the changes in the transition curves. Second, the oxygen
surplus produces no significant doping effect as expected from the simple ionic
model presented above. Therefore, the use of the identity $n = x$ is
well justified. We note, however, that there is a strong oxygen induced
microscopic disorder, which suppresses very effectively superconductivity and
can be detected by magnetization measurements (fishtail effect, see
Fig.~\ref{fig5}(b)). In our well reduced single crystals such irregularities in
the magnetization data have not been observed (see also Fig.~\ref{fig6}).

Since there is only a single impurity site available for oxygen occupation or
reduction in the $214$-compounds, we believe, that the population/depopulation
of this apical O(3) site is closely related to the appearance of
superconductivity. The presence of apical oxygen may cause additional
distortions and strain within the CuO$_2$ plane, preventing the evolution of
the superconducting state independent of the doping $x$. This picture is
consistent with the experimental observations of non-superconducting or weakly
superconducting as-grown crystals depending on the used $p_{\rm O_2}$ during
crystal growth, the evolution/disappearence of superconductivity after
removal/incorporation of additional oxygen as well as with recent transport
measurements on Pr$_{2-x}$Ce$_{x}$CuO$_{4+\delta}$ thin films~\cite{higgins06}.
We do not believe, that the oxygen reduction occurs mainly in the CuO$_2$
planes while the apical site occupation remains unchanged~\cite{richard04}, as
in this case the oxygen variation should trigger a doping effect.

\section{Shubnikov-de Haas Oscillations}
\label{sec:5}

The very high quality of $\rm Nd_{2-x}Ce_{x}CuO_{4+\delta}$ single crystals
obtained following the procedure described above is clearly demonstrated by the
observation of magnetic quantum oscillations in their interlayer resistivity
for doping levels $x = 0.15$, 0.16, and 0.17~\cite{helm09}. A particular
prerequisite for the observation of magnetic quantum oscillations is that the
mean free path $\ell$ of the charge carriers be at least comparable to the
radius $r_B = p_F/eB$ of the cyclotron orbit, where $p_F$ is the Fermi momentum
perpendicular to magnetic field $\mathbf{B}$ and $e$ the elementary charge. For
superconducting cuprates, $r_B$ is a few hundred {\aa}ngstr\"{o}ms in experimentally
available magnetic fields, thus imposing stringent requirements on the crystal
quality. Although $\rm Nd_{2-x}Ce_{x}CuO_{4+\delta}$ is a complex solid 
solution system, this requirement is fulfilled for our high-quality single 
crystals. 

The crystals used in our high magnetic field experiments showed
consistent temperatures and widths of the superconducting
transition for each doping with those presented in Fig. 3(a). The
resistivity ratios, $\rho(273K)/\rho(T_0)$ = 5.6, 5.2, and 9.7,
for $x = 0.15$, 0.16, and 0.17, respectively ($T_0$ is the
superconducting onset temperature), also indicate high crystal
quality.

\begin{figure}[tb]
  \centering
  \includegraphics[width=.95\columnwidth]{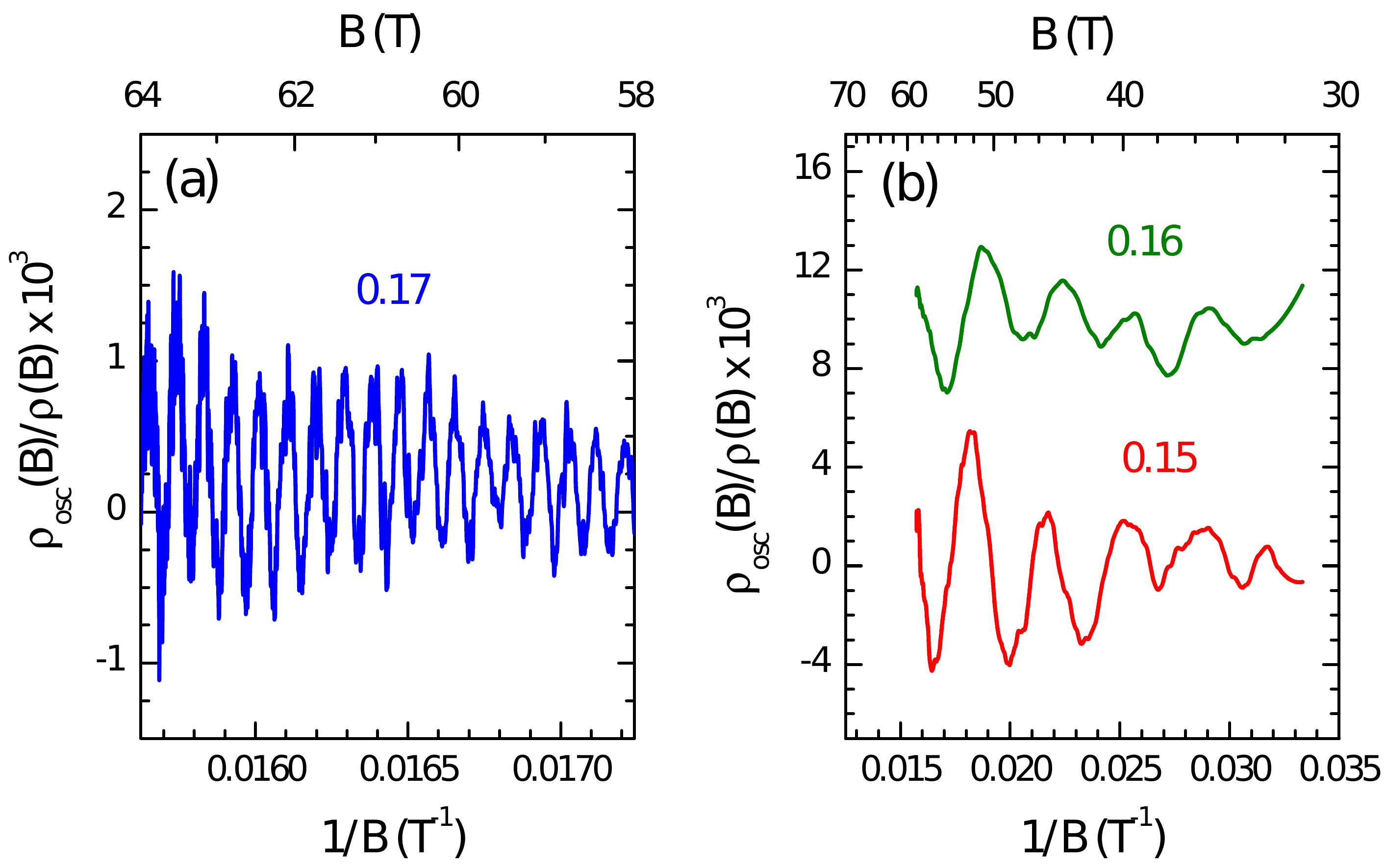}
  \caption{
Shubnikov-de Haas oscillations in $\rm Nd_{2-x}Ce_{x}CuO_{4+\delta}$ 
crystals with (a) $x = 0.17$, and (b) $x
= 0.15$ and 0.16, at a magnetic field applied perpendicular to CuO$_2$ layers.
Note that the oscillation frequency for $x = 0.17$ is 30 times higher than for
the lower doping levels. }
 \label{fig7}
\end{figure}

Fig.~\ref{fig7}(a) shows the oscillatory component of the interlayer
resistivity of the $x = 0.17$ sample at $T = 2.7$ K. The data is obtained from
a raw $R(B)$ curve by subtracting a monotonic background. The oscillations are
periodic in the $1/B$ scale, their positions being independent of
temperature~\cite{helm09}. Such behavior is characteristic of the Shubnikov-de
Haas (SdH) effect originating from the Landau quantization of the electron
spectrum of a metal in a high magnetic field~\cite{shoe84}. The oscillation
frequency, $F_{0.17} = (10950\pm 100)$\,T is directly related to the area of
the cyclotron orbit on the cylindrical Fermi surface: $A_{0.17}= 2\pi
eF_{0.17}/\hbar = (1.04\pm 0.01)\times 10^{20}$~m$^{-2}$. This corresponds to
41.4\% of the first Brillouin zone area, in excellent agreement with the
predictions of band structure calculations~\cite{mass89} and results of
angle-resolved photoemission spectroscopy (ARPES)~\cite{mats07,armi02}.

The samples with lower doping, $x = 0.15$ and 0.16, also show SdH oscillations
(see Fig.~\ref{fig7}(b)), however, at much lower frequencies, $F_{0.15} =
(290\pm 10)$~T and $F_{0.16} = (280\pm 10)$~T, respectively. The corresponding
Fermi surface cross-section amounts to only $\approx 1.1\%$ of the Brillouin
zone area. Such a drastic change of the SdH frequency indicates a
reconstruction of the Fermi surface due to a broken translational symmetry. It
can be explained by introducing a $(\pi/a,\pi/a)$ density-wave
potential~\cite{helm09,chak10}. The slow oscillations are consistent with the
size of small hole pockets formed at the new Brillouin zone boundary as a
result of folding the original large Fermi surface.

Similar results revealing small closed pockets of a reconstructed Fermi surface
have been obtained from quantum oscillation experiments on p-underdoped
YBa$_2$Cu$_3$O$_{6.5}$~\cite{doir07,jaud08,seba08} and
YBa$_2$Cu$_4$O$_8$~\cite{yell08,bang08}. However, by contrast to the latter
compounds, in the n-doped $\rm Nd_{2-x}Ce_{x}CuO_{4+\delta}$  
the ordering potential is manifest already on
the overdoped side of the phase diagram, at $x = 0.16$. Moreover, the
observation of fast oscillations at $x = 0.17$ does not unambiguously rule out
the possibility of the superlattice at this doping. Indeed, our most recent
data on samples with $x = 0.17$~\cite{helm10-} show evidence of
magnetic breakdown, suggesting the Fermi surface to be reconstructed even at
this high doping level. Further work on magnetic quantum oscillations is in
progress, in order to verify this suggestion and obtain more information about
the Fermi surface and its dependence on the carrier concentration in the
electron-doped superconductors.

\section{Summary}

We have grown high quality single crystals of the electron-doped cuprate
superconductors $\rm Nd_{2-x}Ce_{x}CuO_{4+\delta}$ and $\rm
Pr_{2-x}Ce_{x}CuO_{4+\delta}$. We discuss the optimal growth parameters and
annealing conditions as a function of Ce content $x$ required to achieve such
crystals. We show that the oxygen partial pressure of the growth atmosphere has
to be reduce with increasing $x$ in order to avoid microscopic CeO$_2$
precipitations. We also present an optimized annealing process resulting in
narrow transition curves. In the $\rm Nd_{2-x}Ce_{x}CuO_{4+\delta}$ system the
whole phase diagram can be probed with high quality single crystals, whereas in
$\rm Pr_{2-x}Ce_{x}CuO_{4+\delta}$ the overdoped regime is inaccessible due to
the lower solubility limit of Ce. Here, the system
La$_1$Pr$_{1-x}$Ce$_{x}$CuO$_{4+\delta}$ might be an adequate alternative. The
microscopic oxygen distribution can be checked by magnetic measurements. The
as-grown doped crystals exhibit an excess oxygen concentration of
$\delta\approx0.03$, which is nearly completely removed after annealing. The
oxygen reduction is reversible and the removed oxygen comes primary from the
apical site. This apical oxygen governs the evolution of superconductivity by
virtue of induced disorder in the CuO$_2$ planes. The additional oxygen has
only a minor doping effect. The obtained crystals reach a perfection that
quantum oscillations can be observed on such samples.

\section{Acknowledgements}
The authors are very grateful to R. Gross for numerous stimulating and 
fruitful discussions. 
The work was supported the German Research Foundation via the Research Unit FOR~538. 
We also acknowledge support by EuroMagNET II under the EC contract No. 228043.


\begin{thebibliography}{99}

\bibitem{tokura89}Y. Tokura, H. Takagi, S. Uchida, Nature \textbf{337}, 345
    (1989).

\bibitem{Erb} A. Erb, E. Walker, R. Fluekiger, Physica C \textbf{258}, 9
    (1996).

\bibitem{maljuk00} A.N. Maljuk, A.A. Jokhov, I.G. Naumenko, I.K. Bdikin, S.A.
    Zver'kov, G.A. Emel'chenko, Physica C \textbf{329}, 51 (2000).

\bibitem{oka89} K. Oka, H. Unoki, Jpn. J. Appl. Phys. \textbf{28}, L937
    (1989).

\bibitem{kurahashi02}K. Kurahashi, H. Matsushita, M. Fujita, K. Yamada, J.
    Phys. Soc.Jpn.
  \textbf{71}, 910 (2002).

\bibitem{matsuura03}M. Matsuura, P. Dai, H.J. Kang, J.W. Lynn, D.N. Argyriou,
    K. Prokes, Y. Onose,
  Y. Tokura, Phys. Rev. B \textbf{68}, 144503 (2003).

\bibitem{mang04}P.K. Mang, S. Larochelle, A. Mehta, O.P. Vajk, A.S. Erickson,
    L. Lu, W.J.L.
  Buyers, A.F. Marshall, Phys. Rev. B \textbf{70}, 094507 (2004).

\bibitem{uefuji03}T. Uefuji, S. Kuroshima, M. Fujita, and K. Zamada, Physica C
    \textbf{392-396},189 (2003).

\bibitem{kang05}H.J. Kang, P. Dai, H.A. Mook, D.N. Argyriou, V. Sikolenko, J.W.
    Lynn,
  Y. Kurita, S. Seiki, Y. Ando, Phys. Rev. B \textbf{71}, 214512 (2005).

\bibitem{shibata01}H. Shibata, K. Oka, S. Kashiwaya, H. Yamaguchi, Physica C
    \textbf{357}, 363
  (2001).

\bibitem{kim93}J.S. Kim, D.R. Gaskell, Physica C \textbf{209}, 381 (1993).

\bibitem{tallon95}J.L. Tallon, C. Bernhard, H. Shaked, R.L. Hitterman, J.D.
    Jorgensen, Phys. Rev.
  B \textbf{751}, 12911 (1995).

\bibitem{higgins06}J.S. Higgins, Y. Dagan, M.C. Barr, B.D. Weaver, R.L. Greene,
    Phys. Rev. B
  \textbf{73}, 104510 (2006).

\bibitem{radaelli94}P.G. Radaelli, J.D. Jorgensen, A.J. Schultz, J.L. Peng,
    R.L. Greene, Phys. Rev.
  B \textbf{49}, 15322 (1994).

\bibitem{richard04}P. Richard, G. Riou, I. Hetel, S. Jandl, M. Poirier, P.
    Fournier, Phys. Rev. B
  \textbf{70}, 0645131 (2004).

\bibitem{helm09}T. Helm, M. V. Kartsovnik, M. Bartkowiak, N. Bittner,
    M.Lambacher, A. Erb, J. Wosnitza, and Gross, Phys. Rev. Lett. \textbf{103},
    57002 {2009}).

\bibitem{shoe84}D. Shoenberg, {\it Magnetic Oscillations in Metals}
  (Cambridge University Press, Cambridge, 1984).

\bibitem{mass89}S. Massidda, N. Hamada, J. Yu, and A. J. Freeman, Physica C
    \textbf{157}, 571 (1989).

\bibitem{mats07}H. Matsui, T. Takahashi, T. Sato, K. Terashima, H. Ding, T.
    Uefuji, and K. Yamada, Phys. Rev. B \textbf{75}, 224514 (2007).

\bibitem{armi02}P. N. Armitage, F. Ronning, D. H. Lu, C. Kim, A. Damascelli,
    K. M. Shen, D. L. Feng, H. Eisaki, Z.-H. Shen, P.K. Mang et al., Phys. Rev.
    Lett.  \textbf{88}, 257001 (2002).

\bibitem{chak10}J. Eun, X. Jia, and S. Chakravarty, Arxiv:0912.0728;
    unpublished.

\bibitem{doir07}N. Doiron-Leyraud, C. Proust, D. LeBoeuf, J. Levallois, J.-B.
    Bonnemaison, R. Liang, D. A. Bonn, W. N. Hardy, and L Taillefer, Nature
    \textbf{447}, 565 (2007).

\bibitem{jaud08}C. Jaudet, D. Vignolles, A. Audouard, J. Levallois, D.
    LeBoeuf, N. Doiron-Leyraud, B. Vignolle, M. Nardone, A. Zitouni, R. Liang,
    et al., Phys. Rev. Lett. \textbf{100}, 187005 (2008).

\bibitem{seba08}S. E. Sebastian, N. Harrison, E. Palm, T. P. Murphy, C. H.
    Mielke, R. Liang, D. A. Bonn, W. N. Hardy, and G. G. Lonzarich, Nature
    \textbf{454}, 200 (2008).

\bibitem{yell08}E. A. Yelland, J. Singleton, C. H. Mielke, N. Harrison, F. F.
    Balakirev, B. Dabrowski, and J. R. Cooper, Phys. Rev. Lett. \textbf{100},
    047003 (2008).

\bibitem{bang08}A. F. Bangura, J. D. Fletcher, A. Carrington, J. Levallois, M.
    Nardone, B Vignolle, N. Doiron-Leyraud, D. LeBoeuf, L. Taillefer, S. Adachi
    et~al., Phys. Rev. Lett. \textbf{100}, 047004 (2008).

\bibitem{helm10-}T. Helm et al., unpublished.

\end{thebibliography}
\bibliographystyle{epj}

\end{document}